\documentstyle[aps,preprint]{revtex}
\begin{document}
\draft
\preprint{SOGANG-HEP-243/98}
\title{Entropy of 2+1 dimensional de Sitter space in terms of
      brick wall method}
\author{Won Tae Kim\footnote{Electronic address: wtkim@ccs.sogang.ac.kr}}
\address{Department of Physics and Basic Science Research Institute,\\
Sogang University, C.P.O. Box 1142, Seoul 100-611, Korea}
\date{August 1998}
\maketitle
\begin{abstract}
We calculate the statistical entropy of a scalar field on the
background of three-dimensional De Sitter space 
in terms of the brick wall method and finally 
derive the perimeter law of the
entropy.
\end{abstract}
\bigskip

\newpage
Entropy of black holes has an universal area law \cite{bek} and  
the entropy of the well-known Schwarzschild black hole
satisfies the area law by means of thermal radiation based on
the quantum field theory \cite{haw}.
On the other hand, 't Hooft has argued that when one calculates the
black hole entropy, the modes of a quantum field in the vicinity of a 
black hole horizon should be cut off due to gravitational effects
rather than infinitely piling up by suitably choosing a brick-wall cutoff
just beyond the horizon \cite{tho}. Most of brick-wall calculations
have been done for the asymptotically flat cases.

For 2+1 dimensional anti-de Sitter space,
Ba$\tilde{{\rm n}}$ados, Teitelboim, and Zanelli
(BTZ) have obtained a black hole solution which is asymptotically
anti-de Sitter spacetime rather than asymptotically flat \cite{btz}.
The theormodynamic properties has been extensively studied in this black hole
\cite{ct,cal}. Recently the statistical entropy of the de Sitter(DS) space
is studied in terms of Chern-Simons formulation \cite{ms}.
The DS space has a cosmological 
horizon and asymptotically non-flat spacetime, furthermore
the spacetime is bounded by the horizon as the two-dimensional cavity.

In this Brief Report, we shall calculate the entropy of a scalar field
on the DS space background by using the brick wall method. 
As a result, the divergent entropy is obtained
in the vicinity of horizon and by properly choosing the brick-wall cutoff 
we finally obtain the
expected perimeter law. 

Let us start with the following action
\begin{equation}
\label{action}
I=\frac{1}{2\pi} \int d^3 x \sqrt{-g} \left[R -\frac{2}{l^2} \right] ,
\end{equation}
where $\Lambda= \frac{1}{l^2}$ is a cosmological constant.
Then the classical equation of motion yields the DS metric as
\begin{eqnarray}
\label{metric}
ds^2 &=& -g(r) dt^2 +\frac{1}{g(r)} dr^2 + r^2 d \theta^2, \\
g(r) &=& \left(1- \frac{r^2}{l^2} \right). \\
\end{eqnarray}
The horizon is located at $r=l$ and our spacetime is defined within
$0 \le r \le l$. 
The inverse of Hawking temperature is
given by
\begin{equation}
\label{ht}
\beta_H = 2\pi l.
\end{equation}
Let us now introduce a Klein-Gordon field equation on the DS 
background,
\begin{equation}
\label{kg}
\Box \Phi  = 0,
\end{equation}
where we consider the massless case for simplicity.
The Eq. (\ref{kg}) can be solved through the separation of variables
and we can write the wave function as
\begin{equation}
\label{wf}
\Phi(r, \phi, t) = e^{-iEt}e^{i m \phi} R_{Em} (r),
\end{equation}
where $m$ is an azimuthal quantum number of the scalar field.
Then the radial wave equation is written as
\begin{equation}
\label{wave}
\frac{1}{rg(r)} \partial_r [ r g(r) \partial_r R_{Em}(r)] +
k^2 (r,m,E) R_{Em}(r) = 0~,
\end{equation}
where the radial wave number is given by
\begin{equation}
\label{wn}
k^2 (r,m,E) = \frac{1}{g^2(r)} \left[ E^2 - 
            \frac{m^2 g(r)}{r^2} \right]
\end{equation}
in the WKB approximation \cite{tho}.
According to the semi-classical quantization rule, the
radial wave number is quantized as
\begin{equation}
\label{rule}
\pi n_r(m,E) = \int_{L}^{r_H- \epsilon} dr k(r,m,E)~,
\end{equation}
under the brick wall boundary conditions: $\Phi=0$ at $r=L,~r=r_H-\epsilon$. 
Note that $n_r$ is assumed to be a nonnegative integer,
and $\epsilon$ and $L$ are ultraviolet and infrared regulators,
respectively where $\epsilon >0$ and $0 \le L \le r_H -\epsilon$. 
In this range, the energy
$E$ is always positive and the wave number $k$ is real.

The free energy at inverse temperature $\beta$ is given by
\begin{equation}
\label{def}
e^{-\beta F} = \prod_K
                \left[ 1 - e^{-\beta E_K} \right]^{-1}~,
\end{equation}
where $K$ represents the set of quantum numbers.
By using Eq. (\ref{rule}), the free energy can be rewritten as
\begin{eqnarray}
 F &=& \frac{1}{\beta}\sum_K \ln \left[ 1 - e^{-\beta E_K} \right]
   ~\approx ~\frac{1}{\beta} \int dn_r \int dm ~\ln 
            \left[ 1 - e^{-\beta E} \right]
            \nonumber   \\
   &=& -\int dm \int dE ~\frac{n_r}{e^{\beta E} -1}
            \nonumber  \\
\label{free}
   &=& -\frac{1}{\pi} \int dm \int dE
         \frac{1}{e^{\beta E} -1}
         \int dr k(r,m,E)~,
\end{eqnarray}
where we have taken the continuum limit in the first line and integrated
by parts in the second line in Eq. (\ref{free}).
The explicit form of the free energy is given by
\begin{eqnarray}
\label{free00}
 F & =& -\frac{1}{\pi}  \int^\infty_0 dE
         \frac{1}{e^{\beta E} -1}
         \int_L^{r_H-\epsilon} dr \frac{1}{g(r)} \int dm 
           \sqrt{E^2  - \frac{m^2 g(r)}{r^2}} \nonumber \\
   &=& -\frac{1}{2} \int^\infty_0 dE
      \frac{E^2}{e^{\beta E}-1} \int_L^{r_H -\epsilon} dr 
       \frac{r}{g^{3/2}(r)}.   
\end{eqnarray}
Note that the integration with respect to angular variable $m$ is
taken over values for which the square root is real. Performing
the remaining integrations, the free energy is written by  
\begin{equation}
\label{fe}
F  = -\frac{\zeta(3)l^3}{2\beta^3} 
           \left(\frac{1}{\sqrt{l^2-(r_H -\epsilon)^2}}-
                \frac{1}{\sqrt{l^2-L^2}}  \right).
\end{equation}
Let us now evaluate the entropy for the massless
field, which can be obtained from the
free energy (\ref{fe}) at the Hawking temperature,
then the entropy is
\begin{eqnarray}
S &=& \left. \beta^2 \frac{\partial F}{\partial \beta}
              \right|_{\beta=\beta_H}
                                \nonumber \\
\label{entropy0}
  &=& 4\pi a  \left(\frac{l}{\sqrt{l^2-(r_H -\epsilon)^2}}-
                \frac{l}{\sqrt{l^2-L^2}}  \right)
\end{eqnarray}
where the constant is defined by $a \equiv \frac{3 \zeta(3)}{32 \pi^3}$.
This result shows that the entropy behaves as
$1/\sqrt{\epsilon}$ at $\epsilon \rightarrow 0$ which corresponds to
the ultraviolet divergence of the entropy. 
On the other hand, 
the distance of the brick wall from the horizon is related to
the ultraviolet cutoff as
\begin{eqnarray}
\label{invariant}
\tilde{\epsilon}&=&\int_{r_H -\epsilon}^{r_H}
                          \frac{dr}{\sqrt{g(r)}} \\ \nonumber
                &=& l \left(\frac{\pi}{2}
                          -\sin^{-1}\frac{l-\epsilon}{l} 
                     \right).
\end{eqnarray}
Then the entropy (\ref{entropy0})
is neatly represented in terms of
the invariant cutoff (\ref{invariant}) as follows,
\begin{equation}
\label{entropy}
S = 4\pi a \left( \frac{1}{\sin \frac{\tilde{\epsilon}}{l}} -1 \right)
\end{equation}
where we simply fix the infrared cutoff as $L=0$ without loss of
generality since there does not exist any infrared divergence even
though we consider massless scalar field. 
Note that the entropy is always positive in Eq. (\ref{entropy}).
If we choose the cutoff as
\begin{equation}
\tilde{\epsilon}=l \sin^{-1} \left( \frac{a}{a +  l} \right)
\end{equation}
then the entropy is written by the perimeter law
\begin{equation}
\label{area}
S = 2 \cdot 2 \pi r_+.
\end{equation}
Note that for $l \gg a $, the invariant cutoff is simply written as
$\tilde{\epsilon} \approx a$ and it gives the 
entropy $S=4\pi l$.

As a comment, at first sight the infrared regulator $L$ seems to be 
necessary in our massless field, however,
it does not play an important role
in DS space since our spacetime is bounded by the horizon 
and in some sense spacetime is surrounded by the black hole
which is similar to the confined field in the cavity. Therefore,
the finite volume of DS spacetime evaluated as  
$V(r)=2\pi l^2 \left[ 1-\sqrt{ 1-r^2/l^2 } \right]$
removes infrared divergence in contrast to the
asymptotically flat spacetime whose volume is infinite.

\section*{Acknowledgments}
This work was supported by the Korea Research Foundation (1997).



\begin{references}
\bibitem{bek} J. D. Bekenstein, Lett. Nuovo Cimento 4, 737 (1972);
           Phys. Rev. D7, 2333 (1973); D9, 3292 (1974).
\bibitem{haw} S. W. Hawking, Commun. Math. Phys. 43, 199 (1975).
\bibitem{tho} G. 't Hooft, Nucl. Phys. B256, 727 (1985).
\bibitem{btz} M. Ba$\tilde{{\rm n}}$ados, C. Teitelboim, and J. Zanelli,
           Phys. Rev. Lett. 69, 1849 (1992).
\bibitem{ct} S. Carlip and C. Teitelboim, Phys. Rev. D51, 622 (1995).
\bibitem{cal} S. Carlip, Phys. Rev. D51, 632 (1995).
\bibitem{ms} J. Maldacena and A. Strominger, gr-qc/9801096.
\end{references}
\end{document}